\documentclass[prd,twocolumn,showpacs,floatfix,amsmath,amssymb,floatfix]{revtex4}
\usepackage{graphicx,dcolumn,booktabs,bm}
\usepackage{longtable,lscape}
\usepackage{txfonts}
\usepackage{overpic}
\usepackage{multirow}
\usepackage{amssymb}
\usepackage{indentfirst}
\usepackage{feynmf}   
\usepackage{slashed}  
\usepackage{cases}
\usepackage{color,ulem}
\usepackage{graphicx}
\usepackage[colorlinks,
            citecolor=blue,
            anchorcolor=red,
            menucolor=red,
            linkcolor=red,
            filecolor=red,
            runcolor=red,
            urlcolor=blue,
            frenchlinks=red]{hyperref}

\graphicspath{{Figures/}} %

\begin{document}


\title{Strong decay patterns of the hidden-charm pentaquark states $P_c(4380)$ and $P_c(4450)$}
\author{Guang-Juan Wang$^{1}$}\email{wgj@pku.edu.cn}
\author{Li Ma$^{1}$}\email{lima@pku.edu.cn}
\author{Xiang Liu$^{2,3}$}\email{xiangliu@lzu.edu.cn}
\author{Shi-Lin Zhu$^{1,4}$}\email{zhusl@pku.edu.cn}

\affiliation{
$^1$Department of Physics and State Key Laboratory of Nuclear Physics and Technology and Center of High Energy Physics, Peking University, Beijing 100871, China\\
$^2$School of Physical Science and Technology, Lanzhou University, Lanzhou 730000, China \\
$^3$Research Center for Hadron and CSR Physics, Lanzhou University and Institute of Modern Physics of CAS, Lanzhou 730000, China\\
$^4$Collaborative Innovation Center of Quantum Matter, Beijing
100871, China}


\begin{abstract}

With the heavy quark symmetry and spin rearrangement scheme, we
study the strong decay behavior of the hidden-charm pentaquark
states with $J^P={\frac{3}{2}}^{\pm}, {\frac{5}{2}}^{\pm}$ assuming
they are molecular candidates composed of ${\bar D}^{(\ast)}$ and
${\Sigma}_c^{(\ast)}$. We obtain several typical ratios of the
partial decay widths of the hidden-charm pentaquarks. For the three
S-wave $(\bar{D}\Sigma_c^*)$, $(\bar{D}^*\Sigma_c)$,  and
$(\bar{D}^*\Sigma_c^*)$ molecular pentaquarks with
$J^P={{3}/{2}}^{-}$, we have obtained the ratio of their $J/\psi N$ decay
widths: $\Gamma\left[(\bar
D{\Sigma^{\ast}_c})\right]:\Gamma\left[({\bar
D}^{\ast}{\Sigma}_c)\right]:\Gamma\left[({\bar
D}^{\ast}{\Sigma^{\ast}_c})\right]=2.7:1.0:5.4$, which may be useful
to further test the possible molecular assignment of the $P_c$
states.

\end{abstract}

 \maketitle

\section{Introduction}\label{sec1}

Recently, the LHCb Collaboration announced the observation of two
hidden-charm resonances, $P_c(4380)$ and $P_c(4450)$, in the $J/\psi
p$ invariant mass spectrum in the process $\Lambda_b\to J/\psi p K$
\cite{Aaij:2015tga}. The masses and widths of $P_c(4380)$ and
$P_c(4450)$ are\cite{Aaij:2015tga}
\begin{eqnarray*}
M_{P_c(4380)}&=&(4380\pm8\pm29)\, {\rm MeV},\\\Gamma_{P_c(4380)}&=&(205\pm18\pm86)\, {\rm MeV},\\
M_{P_c(4450)}&=&(4449.8\pm1.7\pm2.5)\, {\rm
MeV},\\\Gamma_{P_c(4450)}&=&(39\pm5\pm19)\, {\rm MeV}.
\end{eqnarray*}
Since they are observed in the final state $J/\psi p$, the isospin
of $P_c(4380)$ and $P_c(4450)$ is $I=1/2$. According to LHCb's
analysis, their angular momentum and the parity of the two $Pc$ states are
either $J^P={\frac{3}{2}}^{\pm}$ or ${\frac{5}{2}}^{\pm}$. At
present the spin and parity of each $Pc$ state cannot be determined.

In the literature there exist some theoretical discussions of the
possible hidden-charm pentaquark states
\cite{Wu:2010jy,Yang:2011wz,Burns:2015dwa,Uchino:2015uha,Karliner:2015ina}. Specifically, the possibility of hidden-charm molecular pentaquarks composed
of an anticharmed meson and a charmed baryon was studied
systematically in the framework of the one boson exchange model in Ref.\cite{Yang:2011wz}. In fact, the existence of the hidden-charm
molecular pentaquarks was predicted \cite{Yang:2011wz}.

Let us take the deuteron, which is an extremely loosely
bound molecular state composed of one proton and a neutron with a
binding energy around 2 MeV, as an example. Generally speaking, the binding energy
of the hadronic molecular state is around several to several tens of
MeV. Within the molecular scheme, it is quite natural to understand
the masses of $P_c(4380)$ and $P_c(4450)$, which lie several tens of
MeV below the $(\bar{D}^*\Sigma_c)$ and $(\bar{D}^*\Sigma_c^*)$
threshold. We want to emphasize that the mass difference between
these two $Pc$ states is almost the same as the mass difference
between $\Sigma_c$ and $\Sigma_c^*$, which is around 70 MeV.

Within the molecular scheme, the P-wave, D-wave or even higher
orbital excitations may also exist if the binding energy of the
lowest S-wave hadronic molecule reaches several tens of MeV. For
example, the P-wave state may lie slightly above the S-wave ground
state with an excitation energy around several to tens of MeV. In other
words, the S-wave and P-wave states may completely overlap with each
other. There may exist two or more resonant signals around 4380 MeV
which are close to each other but may carry different parity. If the
P-wave or higher excitation is very broad with a width around 600 MeV, such a state may easily be mistaken as the background. On the
other hand, if the excitation lies several MeV within 4380 MeV but
with a width as narrow as several MeV, then this state may probably
be buried by the $Pc(4380)$ resonance with a width around 205 MeV. The
same situation may also occur around 4450 MeV. The above speculation
may partly explain why the different assignment of the spin and
parity of these two $Pc$ states yields roughly the same good fit \cite{Aaij:2015tga}. The identification of the nearly degenerate
resonances with different parities and widths may require a huge amount of experimental data.

The discovery of $P_c(4380)$ and $P_c(4450)$ opens a new window to
study exotic hadronic matter. The recent discovery of $P_c(4380)$ and $P_c(4450)$ has inspired theorists'
extensive interest in these two states. With the one pion exchange
model, Chen, Liu, Li, and Zhu performed a dynamical calculation of
the $\Sigma_c \bar{D}^*$ and $\Sigma_c^*\bar{D}^*$ systems. The
results confirm that there do exist two S-wave $\Sigma_c \bar{D}^*$
and $\Sigma_c^*\bar{D}^*$ molecular states around the mass regions of
$P_c(4380)$ and $P_c(4450)$, respectively \cite{Chen:2015loa}.

More recently, the authors of Ref .\cite{Chen:2015moa} constructed the
spin-3/2 and spin-5/2 hidden-charm local pentaquark interpolating
currents to investigate $P_c(4380)$ and $P_c(4450)$ within the
framework of the QCD sum rule formalism. The spectra of the two newly observed $P_c$ states can be reproduced well, and two extra
hidden-charm pentaquarks and two hidden-bottom pentaquarks are
predicted. Unfortunately, these currents do not distinguish a
tightly bound pentaquark structure or a molecular structure composed
of an anticharmed meson and a charmed baryon.

In Ref. \cite{Mironov:2015ica}, Mironov and Morozov analyzed four
possibilities of the configuration of pentaquarks qualitatively.
They claimed that the internal color components of the pentaquark
may play a crucial role in forming the two $P_c$ states \cite{Mironov:2015ica}. The Bethe-Saltpeter equation was applied to
studying the interactions of $\bar{D}\Sigma_c^*$ and $\bar D^*\Sigma$,
where $P_c(4380)$ and $P_c(4450)$ were explained as the
$\bar{D}\Sigma_c^*$ state with $J^P=3/2^-$ and $\bar D^*\Sigma$ with
$J^P=5/2^+$, respectively \cite{He:2015cea}.

Lebed investigated the hidden-charm pentaquark through the dynamical
diquark picture carefully \cite{Lebed:2015tna}, where the two $P_c$
states were composed of color-antitriplet diquark $cu$ and
color-triplet triquark $\bar{c}(ud)$. In Ref. \cite{Maiani:2015vwa}, the authors suggest that the two $Pc$ states have a configuration of diquark-diquark-antiquark. The total spin of the light diquark and the orbital excitation in the pentaquark states combine to explain the mass difference. In Ref. \cite{Wang:2015epa},
the diquark-diquark-antiquark-type interpolating currents were
introduced to study the two $P_c$ states using QCD sum rules. The
same formalism was extended to study the hidden-charm pentaquark
with $J^P=1/2^\pm$ in Ref. \cite{Wang:2015ava}.

With the bound state version of the topological soliton model for
baryons, Scoccola, Riska, and Rho noticed the existence of a bound
(or quasibound) $\bar{D}$-soliton state which is compatible with
the two $P_c$ states \cite{Scoccola:2015nia}. The quark
delocalization color screening model was adopted to study the
hidden-charm molecular pentaquarks, where $P_c(4380)$ is suggested
to be a mixed structure of $\Lambda_c\bar D^*$, $\Sigma_c \bar D^*$,
$\Sigma_c^* \bar D$, and $\Sigma_c^*\bar D^*$ with
$I(J^P)=1/2(3/2^-)$, while $P_c(4450)$ can be a $\Sigma_c^*\bar D^*$
state with $I(J^P)=1/2(5/2^-)$ \cite{Huang:2015uda}. Zhu and Qiao
used a constituent diquark-triquark model to explain the two $P_c$
states in Ref. \cite{Zhu:2015bba}. $P_c(4450)$ was proposed as a $\chi_{c1}p$ resonance in Ref. \cite{Meissner:2015mza}.

Besides the mass spectrum, the productions of the hidden-charm
pentaquarks were investigated in the weak decays of the bottom
baryons in the SU(3) limit \cite{Li:2015gta,Cheng:2015cca}, through
the photoproductions
\cite{Wang:2015jsa,Kubarovsky:2015aaa,Karliner:2015voa} and the $\pi^-p
\to J/\psi n$ reaction \cite{Lu:2015fva}.

Before closing the brief review of the present research status of
these two $P_c$ states, we need to mention that there also exist
several nonresonant explanations of the $Pc$ signals. In Ref. \cite{Guo:2015umn}, $P_c(4450)$ was explained as the kinematical
effect due to the rescattering process $\chi_{c1}p$ to $J/\psi p$.
It was pointed out that $P_c(4380)$ and $P_c(445)$ can also be
understood by the triangle singularity \cite{Liu:2015fea}.

Although there were some theoretical studies of $P_c(4380)$ and
$P_c(4450)$, as discussed above, more efforts are demanded to reveal
the underlying properties of two $P_c$ states and to distinguish
different theoretical assignments of these two $Pc$ states. In particular, the decay behaviors of $P_c(4380)$ and $P_c(445)$ can
provide us with useful information about their inner structures.

In this work, we will focus on the strong decay patterns of
$P_c(4380)$ and $P_c(4450)$ assuming that they are molecular states.
Since the D-wave or higher partial waves are strongly suppressed by
phase space, we focus on the S-wave and P-wave decay of these
states. We adopt the spin rearrangement scheme in the heavy quark
limit, which was applied to study the strong decays of $Z_1(4475)$
\cite{Ma:2014zua}. With the same formalism, a comprehensive
investigation of the decay pattern and production mechanism of $XYZ$
states was performed in Ref. \cite{Ma:2014zva}.

We will calculate the ratios of the strong decay widths of
$P_c(4380)$ and $P_c(4450)$ with spin-parity $J^{P}={{3}/{2}}^{\pm},
{{5}/{2}}^{\pm}$ under the molecular assignments. These predicted
ratios of $P_c(4380)$ and $P_c(4450)$ can be measured at future
experiments, which are useful for testing the $\bar
D^{(\ast)}{\Sigma}^{(\ast)}_c$ molecular assignments of $P_c(4380)$
and $P_c(4450)$.

This paper is organized as follows. We
decompose the initial and final states into the heavy spin and light
spin in the heavy quark limit in Sec. \ref{sec2}. We give the
general expressions for the $T$ matrix element of the decay process.
In Sec. \ref{sec3}, we present the numerical results. In Sec. \ref{sec4}, we discuss our results and conclude.

\section{The spin rearrangement scheme}
\label{sec2}

As an approximate symmetry, heavy quark symmetry is applied to study
the structures of hadrons which contain heavy quarks. In the heavy
quark limit, the total angular momentum $J$ of a system can be
decomposed into two parts, i.e., the heavy spin $S_H$ and light spin
$S_l$, which satisfy the relation $\vec{S}_l+\vec{S}_H=\vec{J}$.
Here, the light spin denotes the light degrees of freedom including
all of the orbital angular momenta and the spin of light quarks within a
hadron, while the heavy spin is the total spin of the heavy quarks.

In general, under the spin arrangement scheme, the hidden-charm
molecular pentaquark states composed of an anticharmed meson
$\bar{D}^{(*)}$ and a charmed baryon ${Q}_c^{(*)}$, which have the
total angular momentum $J$, can be decomposed as
\begin{widetext}
\begin{eqnarray}
\left |\bar{D}^{(\ast)}{{{Q}_c}^{{(\ast)}}}\right\rangle&=&\left[\left[[\bar{c}\otimes({q_1}\otimes L)_s]_g\otimes[c\otimes ({{q_2}{q_3}})_m]_k\right]_{J_0}\otimes L' \right]_J    |(\bar{c}{q_1})(c{q_2}{q_3})\rangle  \nonumber\\
 &=&\sum_{h=0}^1 \sum_{n={|R-L|}}^{R+L} \sum_{R=|m-\frac{1}{2}|}^{m+\frac{1}{2}} \sum_{T=|n-L'|}^{n+L'} (-1)^{L+m+s+R+h+n+L'+J}\hat{s}\hat{R}\hat{h}\hat{n}\hat{g}\hat{k}\hat{J_0}\hat{T}\nonumber\\
 &&\times \left \{
        \begin{array}{c c c}
          1/2   &    s   &  g  \\
          1/2   &    m   &  k  \\
          h   &    n   & J_0
      \end{array}
\right \}
 \left \{
        \begin{array}{c c c}
          L   &    1/2   &  s  \\
          m   &    n   &  R
      \end{array}
\right \}\left \{
        \begin{array}{c c c}
          h   &    n   &  J_0  \\
          L'   &    J   &  T
      \end{array}
\right \} \Bigg|\left[\left[\left[\bar{c}c\right]_h\otimes\left[(\left[{q_1}{(q_2{q_3})_m}\right]_R\otimes L\right)_n\otimes L'\right]_T\right]_J
 \Bigg\rangle |(\bar{c}c)(q_1{q_2}{q_3})\rangle \nonumber\\
&=&
\sum{\mathcal{G}^{m,L,n,L',J}_{h,R,T}}\left|\left[\left[\left[\bar{c}c\right]_h\otimes\left[(\left[{q_1}{(q_2{q_3})_m}\right]_R\otimes
L\right)_n\otimes L'\right]_T\right]_J\right\rangle \left
|(\bar{c}c)(q_1{q_2}{q_3})\right\rangle\label{ww1},
\end{eqnarray}
\end{widetext}
with $\hat{J}=\sqrt{2J+1}$. Here, $L$ denotes the orbital angular
momentum within the anticharmed meson, while $L'$ is the orbital
angular momentum between the anticharmed meson and the charmed baryon.
$s$ and $g$ represent the light spin and the total angular momentum
of the anticharmed meson, respectively. $m$ and $k$ are the light
spin and the total angular momentum of the charmed baryon,
respectively. $J_0$ denotes the total angular momentum of the
anticharmed meson and the charmed baryon. $J$ is the angular momentum
of the pentaquark state. For the S-wave molecular state, $J=J_0$.

We also define $h$ and $T$ as the heavy spin and total light spin of
the system. $R$ is the total spin of the three light quarks. We
denote the sum of the spin $R$ and the orbital angular momentum $L$
within the anticharmed meson as the angular momentum $n$. In
addition, $|(\bar{c}{q_1})(c{q_2}{q_3})\rangle$ and
$|(\bar{c}c)(q_1{q_2}{q_3})\rangle$ denote the flavor wave
functions, where the general expressions $|(ab)\rangle$ and
$|(abc)\rangle$ are the abbreviations of $|ab+ba\rangle/\sqrt{2}$
and $|abc+bac+acb+cab+cba+bca\rangle/\sqrt{6}$, respectively.

In the following, we focus on the S-wave $\bar{D}^*\Lambda_c$,
$\bar{D}^{(*)}\Sigma_c^{(*)}$ molecular states with $J^P=3/2^-$ and
$J^P=5/2^-$, and the P-wave $\bar{D}^{(*)}\Lambda_c$ and $\bar{D}^{(*)}\Sigma_c^{(*)}$ molecular states with $J^P=3/2^+$
and $J^P=5/2^+$. According to Eq. (\ref{ww1}), we can perform the
decomposition of the initial hidden-charm molecular pentaquarks,
where the relevant terms and the corresponding coefficients are
listed in Table \ref{initial state}.

\renewcommand{\arraystretch}{1.6}
\begin{table*}[htbp]
 \caption{The decomposition of the hidden-charm molecular pentaquarks.
We list the coefficients ${\mathcal{G}^{m,L,n,L',J}_{h,R,T}}$ in Eq. (\ref{ww1}) for the combination $[h,R,T]$, which stands for
$\left|\left[\left[\left[\bar{c}c\right]_h\otimes\left[(\left[{q_1}{(q_2{q_3})_m}\right]_R\otimes
L\right)_n\otimes L'\right]_T\right]_J\right\rangle$. We use the
subscripts $H$ and $l$ to label the heavy and light spins,
respectively. The notation $\cdots$ indicates that this combination is
forbidden for the S-wave and P-wave molecular pentaquarks. We use 0 to denote the
combination which is suppressed by the heavy quark
symmetry.}\label{initial state}
\begin{center}
   \begin{tabular}{c|c c ccccccc} \toprule[1pt]
     &\multicolumn{3}{c}{${J^P={\frac{3}{2}}^{-}}$}  & &\multicolumn{1}{c}{${J^P={\frac{5}{2}}^{-}}$}&\\

    & $\left[0_H,{\frac{3}{2}},{\frac{3}{2}}_l\right]$
      & $\left[1_H,{\frac{1}{2}},{\frac{1}{2}}_l\right]$  &  $\left[1_H,{\frac{3}{2}},{\frac{3}{2}}_l\right]$ &  &$\left[1_H,{\frac{3}{2}},{\frac{3}{2}}_l\right]$ \\\midrule[1pt]
  $\left|\bar{D}^{\ast}{{\Lambda}_c}\right\rangle$ & $0$ &$1$
      &$0$ & & $\cdots$\\
      $\left|\bar{D}{\Sigma_c^{\ast}}\right\rangle$ & $\frac{1}{2}$ &$-\frac{1}{\sqrt{3}}$
      &$\frac{\sqrt{5}}{2\sqrt{3}}$ &  &$\cdots$\\
  $\left|\bar{D}^{\ast}{\Sigma}_c\right\rangle$ & $-\frac{1}{\sqrt{3}}$ & $\frac{1}{3}$ &  $\frac{\sqrt{5}}{3}$ &&$\cdots$\\

      $|\bar{D}^{\ast}{{\Sigma}_c^{\ast}}\rangle$ & $\frac{\sqrt{5}}{2\sqrt{3}}$ &$\frac{\sqrt{5}}{3}$       &$\frac{1}{6}$  & & $1$\\

 \bottomrule[1pt]

     &\multicolumn{7}{c}{${J^P={\frac{3}{2}}^{+}}$}   &&\\

     & $\left[0_H,\frac{1}{2},{\frac{3}{2}}_l\right]$& $\left[0_H,\frac{3}{2},{\frac{3}{2}}_l\right]$
      &  $\left[1_H,\frac{1}{2},{\frac{1}{2}}_l\right]$ & $\left[1_H,\frac{3}{2},{\frac{1}{2}}_l\right]$  &  $\left[1_H,\frac{1}{2},{\frac{3}{2}}_l\right]$ & $\left[1_H,\frac{3}{2},{\frac{3}{2}}_l\right]$      & $\left[1_H,\frac{3}{2},{\frac{5}{2}}_l\right]$ \\\midrule[1pt]

     $\left|\bar{D}^{\ast}{\Lambda}_c\right\rangle(J_0=\frac{1}{2})$ & $\frac{\sqrt{3}}{2}$ &0 &$\frac{{1}}{3}$ & 0 &$-\frac{\sqrt{5}}{6}$  &0 &0 \\
     $\left|\bar{D}^{\ast}{\Lambda}_c\right\rangle(J_0=\frac{3}{2})$ & 0 & 0 &$\frac{\sqrt{5}}{3}$ & 0 &$\frac{2}{3}$  &0 &0 \\

    $\left|\bar{D}{\Sigma_c^{\ast}}\right\rangle(J_0=\frac{3}{2})$ &0 & $\frac{1}{2}$ &$-\frac{1}{3}{\sqrt{\frac{5}{3}}}$ & $-\frac{1}{3}{\sqrt{\frac{5}{6}}}$ &$-\frac{2}{3\sqrt{3}}$ &$\frac{1}{6}\frac{11}{\sqrt{15}}$ &$\frac{1}{\sqrt{10}}$\\

  $\left|\bar{D}^{\ast}{\Sigma_c}\right\rangle(J_0=\frac{1}{2})$ &$-\frac{1}{2\sqrt{3}}$ &0 &$-\frac{5}{9}$ & $\frac{1}{9\sqrt{2}}$ &  $\frac{5\sqrt{5}}{18}$ &$-\frac{2}{9}$  &$\frac{1}{\sqrt{6}}$\\
  $\left|\bar{D}^{\ast}{\Sigma_c}\right\rangle(J_0=\frac{3}{2})$ &0 &$-\frac{1}{\sqrt{3}}$  &$\frac{\sqrt{5}}{9}$ & $-\frac{\sqrt{10}}{9}$ &  $\frac{2}{9}$ &$\frac{11}{9\sqrt{5}}$  &$\sqrt{\frac{2}{15}}$\\

  $\left|\bar{D^{\ast}}{\Sigma_c^{\ast}}\right\rangle(J_0=\frac{1}{2})$ & $\sqrt{\frac{2}{3}}$ &0 &$-\frac{2\sqrt{2}}{9}$       &$-\frac{1}{18}$  &  $\frac{\sqrt{10}}{9}$ &  $\frac{\sqrt{2}}{9}$ &  $-\frac{1}{2\sqrt{3}}$\\

  $\left|\bar{D^{\ast}}{\Sigma_c^{\ast}}\right\rangle(J_0=\frac{3}{2})$ &0 &$\frac{1}{2}\sqrt{\frac{5}{3}}$ &$\frac{5}{9}$       &$-\frac{1}{9\sqrt{2}}$  &  $\frac{2\sqrt{5}}{9}$ &  $\frac{11}{90}$ &  $\frac{1}{5\sqrt{6}}$\\

  $\left|\bar{D^{\ast}}{\Sigma_c^{\ast}}\right\rangle(J_0=\frac{5}{2})$ & 0 &0  &0 &$\frac{\sqrt{3}}{2}$ &0    &$\frac{\sqrt{6}}{5}$  &  $\frac{1}{10}$\\

 \bottomrule[1pt]

     &\multicolumn{4}{c}{${J^P={\frac{5}{2}}^{+}}$}   &&\\

    & $\left[0_H,\frac{3}{2},{\frac{5}{2}}_l\right]$
      &  $\left[1_H,\frac{1}{2},{\frac{3}{2}}_l\right]$ & $\left[1_H,\frac{3}{2},{\frac{3}{2}}_l\right]$  & $\left[1_H,\frac{3}{2},{\frac{5}{2}}_l\right]$ \\\midrule[1pt]
  $\left|{\bar{D}}^{\ast}{\Lambda_c}\right\rangle$ & $0$ &$1$ & $0$ &$0$  \\
      $\left|\bar{D}{\Sigma^{\ast}}\right\rangle$ & $\frac{1}{2}$ &$-\frac{1}{\sqrt{3}}$ & $-\frac{1}{\sqrt{15}}$ &$\frac{1}{2}{\sqrt{\frac{7}{5}}}$  \\

  $\left|\bar{D}^{\ast}{\Sigma_c}\right\rangle$ & $-\frac{1}{\sqrt{3}}$ & $\frac{1}{3}$ &  $-\frac{2}{3\sqrt{5}}$ &$\sqrt{\frac{7}{15}}$\\

  $\left|\bar{D^{\ast}}{\Sigma_c^{\ast}}\right\rangle(J_0=\frac{3}{2})$ & $\frac{\sqrt{5}}{2\sqrt{3}}$ &$\frac{\sqrt{5}}{3}$       &$-\frac{1}{15}$  &  $\frac{1}{10}{\sqrt{\frac{7}{3}}}$\\
  $\left|\bar{D^{\ast}}{\Sigma_c^{\ast}}\right\rangle(J_0=\frac{5}{2})$ & 0 &0       &$\frac{\sqrt{21}}{5}$  &  $\frac{2}{5}$\\

 \bottomrule[1pt]
      \end{tabular}
  \end{center}
\end{table*}

We make the same decomposition of the final states when these
hidden-charm pentaquarks decay into a charmonium plus a baryon. If
there exists an $L''$ excitation between a charmonium and a baryon,
the final state can generally be written as follows under the spin
rearrangement scheme:
\begin{eqnarray}
 && \left|\mathrm{Charmonium}\right\rangle\otimes\left|\mathrm{Baryon} \right\rangle\nonumber
\\& &= \left[[(\bar{c}c)_{g'}\otimes L ]_{k'}\otimes [Q\otimes L'' ]_{J'_0}\right]_J |(\bar{c}c)\rangle|({q_1{q_2}{q_3}})\rangle  \nonumber\nonumber\\
  &&=\sum_{T=|L-{J'_0}|}^{L+{J'_0}} (-1)^{J'_0+L+g'+J}\sqrt{2T+1}\sqrt{2k'+1}\left \{
        \begin{array}{c c c}
          {J'_0}   &    L   &  T  \\
          g'   &    J   &  k'
      \end{array}
\right \} \nonumber\\
  &&\quad\times \Bigg|\left[\left(\bar{c}c\right)_{g'}\otimes\left[L\otimes {J'_0}\right]_T\right]_J \Bigg\rangle |(\bar{c}c)\rangle|({q_1}{q_2}{q_3})\rangle\nonumber\\
&&=\sum{\mathcal{F}^{g',L,k',L'',J'_0,J}_{g',T}}\left|\left[\left(\bar{c}c\right)_{g'}\otimes\left[L\otimes {J'_0}\right]_T\right]_J \right\rangle\left |(\bar{c}c)\right\rangle\left|({q_1}{q_2}{q_3})\right\rangle,\label{w2}\nonumber\\
   \end{eqnarray}
where $g'$, $L$ and $k'$ are the spin, orbital, and angular momentum
quantum numbers of the charmonium, respectively. $Q$ stands for the
total angular momentum of the baryon. The coupling of $Q$ and $L''$
forms $J'_0$. $T$ and $g'$ are the light spin and heavy spin,
respectively. We list the kinematically allowed final states in
Table \ref{final state} when the hidden-charm pentaquarks lie in the
mass range of $4430\sim4450$ MeV.

For a molecular state with the configuration $\bar{D}^{(*)}
Q_c^{(*)}$, $J_0$ is the sum of the spins of the meson and baryon,
which further couples with the orbital angular momentum $L$ to form
the total angular momentum $J$ of the pentaquark. For a fixed $J$,
there exist different combinations of $J_0$ and $L$ if $L$ is
nonzero. In Table \ref{initial state}, we mark the corresponding
$J_0$ values for some P-wave $\bar{D}^{(*)} Q_c^{(*)}$ molecular
states. In a similar way, we can also deal with the coupling of the
angular momentum of the final states. For an initial
$\bar{D}^{(\ast)}{{Q^{(\ast)}_c}}$ molecular state, we have
\begin{eqnarray}
  |\bar{D}^{(\ast)}{{Q^{(\ast)}_c}}\rangle&=&{{\bar{D}^{(\ast){g_m}}_g}}\otimes{{Q^{(\ast)}_k}^{k_m}}\otimes{L^{L_m}}\nonumber\\
  &=&\sum_{J} \sum_{J_0}\left\langle J_0(g_m+k_m)|g{g_m}k{k_{m}}\right \rangle \nonumber\\&&\times\left\langle J(g_m+k_m+L_m)|J_0(g_m+k_m)L{L_{m}}\right\rangle \nonumber\\&&\times\left|\left[\left[{\bar{D}_g}^{(\ast)} {Q_k}^{(\ast)}\right]_{J_0}\otimes L\right]^{J}_{J_m} \right\rangle,
  \label{w3}
\end{eqnarray}
where we adopt the same notations as in Eq. (\ref{ww1}). For the
final states, we have
\begin{eqnarray}
&& |\mathrm{charmonium}\rangle \otimes |N\rangle
\nonumber\\&&={[{(c\bar c)_{g'}} \otimes L]^{k'_m}_{k'}}\otimes{{Q^{Q_m}}}\otimes{L'^{L'_m}}\nonumber\\
  &&=\sum_{J'} \sum_{J'_0} \left\langle J'_0(L'_m+Q_m)|QQ_m L'{L'_{m}}\right\rangle\nonumber\\&&\quad\times\left\langle J'(k'_m+Q_m+L'_m)|k'{k'_{m}}J'_0(L'_m+Q_m)\right\rangle \nonumber\\
&&\quad\times \left|\left[ \left[ {(c\bar c)_{g'}]}\otimes L\right]_{k'} \otimes\left[ Q\otimes
L'\right]_{J'_0}\right]^{J'}_{J'_M} \right \rangle, \label{w4}
\end{eqnarray}
where $N$ represents the nucleon. The notations are the same as in
Eq. (\ref{w2}).

\renewcommand{\arraystretch}{1.9}
\begin{table*}[htbp]
\caption{The coefficients $\mathcal{F}^{g',L,k',L'',J'_0,J}_{g',T}$
in Eq. (\ref{w2}) corresponding to different combinations of $[g,T]$, which is the abbreviation of
$\Bigg|\left[\left(\bar{c}c\right)_{g'}\otimes\left[L\otimes
{J'_0}\right]_T\right]_J \Bigg\rangle$. We use $\cdots$ to mark the
forbidden combination for the S-wave and P-wave decays, while we use
0 to denote the combination which is suppressed by the heavy quark
symmetry.}\label{final state}
\begin{center}
{ \small
   \begin{tabular}{c|cccccccccccccccc} \toprule[1pt]
     & \multicolumn{1}{c}{${J={\frac{3}{2}}^{-}}$}&  &\multicolumn{3}{c}{${J={\frac{3}{2}}^{+}}$} &  &\multicolumn{1}{c}{${J={\frac{5}{2}}^{+}}$}&\\

    & $\left[1_H,{\frac{1}{2}}_l\right]$
      & &$\left[1_H,{\frac{1}{2}}_l\right]$  &  $\left[1_H,{\frac{3}{2}}_l\right]$ &  $\left[0_H,{\frac{3}{2}}_l\right]$  & &$\left[1_H,{\frac{5}{2}}_l\right]$  \\\midrule[1pt]

      $\left|J/{\psi}N\right\rangle$ & 1& &$\cdots$
      & $\cdots$ & $\cdots$ && $\cdots$ \\

       $|\chi_{c0}(1^3P_0)N\rangle$ & $\cdots$ && $\cdots$
       & $\cdots$ & $\cdots$ && $\cdots$ \\

     $\left|\chi_{c1}(1^3P_1)N\right\rangle$ & $\cdots$ && $-\frac{1}{\sqrt{6}}$
       & $\frac{\sqrt{5}}{\sqrt{6}}$ & 0 && $\cdots$\\

      $\left|\chi_{c2}(1^3P_2){N}\right\rangle$ &$\cdots$&&${\frac{\sqrt{5}}{\sqrt{6}}}$
       &  $\frac{1}{\sqrt{6}}$ & 0 & &1 \\
      $\left|h_c(1^1P_1)N\right\rangle$ & $\cdots$ && 0
       & 0 & 1&& 0 \\

      $\left|\eta_c(1^1S_0)N\right\rangle$ &$\cdots$&&$\cdots$
      &$\cdots$&$\cdots$&&$\cdots$\\

      \bottomrule[1pt]

      &\multicolumn{3}{c}{${J={\frac{3}{2}}^{+}}$}   &&\multicolumn{1}{c}{${J={\frac{5}{2}}^{+}}$}  &&\multicolumn{4}{c}{${J={\frac{3}{2}}^{-}}$} &&\multicolumn{3}{c}{${J={\frac{5}{2}}^{-}}$}\\\midrule[1pt]

   & $\left[0_H,{\frac{3}{2}}_l\right]$ & $\left[1_H,{\frac{1}{2}}_l\right]$
      & $\left[1_H,{\frac{3}{2}}_l\right]$  &&  $\left[1_H,{\frac{3}{2}}_l\right]$  && $\left[0_H,{\frac{3}{2}}_l\right]$
    & $\left[1_H,{\frac{1}{2}}_l\right]$
      & $\left[1_H,{\frac{3}{2}}_l\right]$  &  $\left[1_H,{\frac{5}{2}}_l\right]$  && $\left[0_H,{\frac{5}{2}}_l\right]$ & $\left[1_H,{\frac{3}{2}}_l\right]$  &  $\left[1_H,{\frac{5}{2}}_l\right]$ \\\midrule[1pt]

      $|J/{\psi}N\rangle(J_0=\frac{1}{2})$ & 0 & 1 &0&
      & $\cdots$ && $\cdots$ & $\cdots$ & $\cdots$ & $\cdots$& & $\cdots$  & $\cdots$ & $\cdots$\\

      $|J/{\psi}N\rangle(J_0=\frac{3}{2})$  & 0 &0
      & 1  && 1 && $\cdots$ & $\cdots$& $\cdots$ & $\cdots$& &$\cdots$& $\cdots$ & $\cdots$\\

       $|\chi_{c0}(1^3P_0)N\rangle(J_0=\frac{3}{2})$  & $\cdots$ & $\cdots$ & $\cdots$&
       & $\cdots$&& 0 & $\frac{1}{\sqrt{6}}$ &$-\frac{1}{\sqrt{3}}$ & $\frac{1}{\sqrt{2}}$ && $\cdots$ & $\cdots$& $\cdots$\\

     $|\chi_{c1}(1^3P_1)N\rangle(J_0=\frac{1}{2})$   & $\cdots$ & $\cdots$ & $\cdots$ && $\cdots$  &&0 & $-\frac{1}{\sqrt{6}}$ & $\frac{\sqrt{5}}{\sqrt{6}}$
       & 0 && $\cdots$& $\cdots$& $\cdots$\\

 $|\chi_{c1}(1^3P_1)N\rangle(J_0=\frac{3}{2})$  & $\cdots$ & $\cdots$ & $\cdots$ && $\cdots$ &&0 & $-\frac{\sqrt{5}}{2\sqrt{3}}$ & $\sqrt{\frac{2}{15}}$
       & $\frac{3}{2\sqrt{5}}$ &&0 &$-\sqrt{\frac{3}{10}}$ & $\sqrt{\frac{7}{10}}$\\

$|\chi_{c2}(1^3P_2){N}\rangle(J_0=\frac{1}{2})$   & $\cdots$ & $\cdots$ & $\cdots$ &&
$\cdots$ &&0 &${\frac{\sqrt{5}}{\sqrt{6}}}$
       &  $\frac{1}{\sqrt{6}}$ &0 &&0 &1  &0  \\

    $|\chi_{c2}(1^3P_2){N}\rangle(J_0=\frac{3}{2})$  & $\cdots$ & $\cdots$ & $\cdots$ && $\cdots$ &&0 &$\frac{\sqrt{5}}{2\sqrt{3}}$
       &  $2\sqrt{\frac{2}{15}}$ &$\frac{1}{2\sqrt{5}}$ &&0 &$\sqrt{\frac{7}{10}}$ & $\sqrt{\frac{3}{10}}$  \\

   $|\eta_c(1^1S_0)N(J_0=\frac{3}{2})\rangle$ &1 &0&0&
      &$\cdots$&&$\cdots$&$\cdots$&$\cdots$&$\cdots$&&$\cdots$&$\cdots$&$\cdots$\\

      $|h_c(1^1P_1)N\rangle(J_0=\frac{1}{2})$  & $\cdots$
       & $\cdots$ &$\cdots$ &&$\cdots$& &1 &0 & 0&0&& $\cdots$& $\cdots$ &$\cdots$\\
 $|h_c(1^1P_1)N\rangle(J_0=\frac{3}{2})$ & $\cdots$
       & $\cdots$ &$\cdots$ &&$\cdots$& &1 &0 & 0&0&& 1& 0 &0\\

      \bottomrule[1pt]
      \end{tabular}}
  \end{center}
\end{table*}

For a hidden-charm pentaquark decay into the charmonium and nucleon
\begin{eqnarray*}
\bar{D}^{(\ast)}{{Q^{(\ast)}_c}}\to {\rm{Charmonium}}+{N},
\end{eqnarray*}
its $T$ matrix element reads
\begin{widetext}
\begin{eqnarray}
|T|^2 &=&A\sum_{g_m,k_m,{L_m}} \sum_{k'_m,Q_m,{L'_m}}\left|\left\langle \left[\mathrm{Charmonium}\otimes N\right]^{J}\left|H\right|\left[\bar{D}^{(\ast)}{{{\Sigma}_c}^{(\ast)}}\right]^{J}\right\rangle\right|^2\nonumber\\
&=&A\sum_{g_m,k_m,{L_m}} \sum_{k'_m,Q_m,{L'_m}} \sum_{J_0,{J_1}_0} \sum_{J'_0,{J'_1}_0}{{\left\langle J'_0(k'_m+Q_m)|k'{k'_{m}},QQ_m\right\rangle}^{\ast}} {\left\langle J{J_m}|J'_0(k'_m+Q_m)L'{L'_{m}}\right\rangle}^{\ast} \left\langle J_0(g_m+k_m)|g{g_m}k{k_{m}}\right\rangle  \nonumber\\
&& \times  {\left\langle J(g_m+k_m+L_m)|J_0(g_m+k_m)L{L_{m}}\right\rangle} {{\left\langle {J'_1}_0(k'_m+Q_m)|k'{k'_{m}},QQ_m\right\rangle} {\left\langle J{J_m}|{J'_1}_0(k'_m+Q_m)L'{L'_{m}}\right\rangle}}   \nonumber\\
&& \times {{\left\langle {{J_1}_0}(g_m+k_m)|g{g_m}k{k_{m}}\right\rangle}^{\ast} {\left\langle J(g_m+k_m+L_m)|{{J_1}_0}(g_m+k_m)L{L_{m}}\right\rangle}^{\ast}} {\delta}\left({g_m}+{k_m}+{L_m}-{k'_m}-{Q_m}-{L'_m}\right)\nonumber\\
&&\times\left \langle {\left[\left[{(c\bar c)_{g'}\otimes L}\right]_{k'} \otimes \left[Q\otimes
L'\right]_{J'_0}\right]^{J}}{\parallel}H{\parallel}{\left[\left[{\bar{D}_g}^{(\ast)}
{\Sigma_k}^{(\ast)}\right]_{J_0}\otimes L\right]^{J}} \right\rangle
{\left\langle {\left[ \left[{(c\bar c)_{g'}\otimes L}\right]_{k'} \otimes \left[Q\otimes
L'\right]_{{J'_1}_0}\right]^{J}}{\parallel}H{\parallel}{\left[\left[{\bar{D}_g}^{(\ast)}
{\Sigma_k}^{(\ast)}\right]_{{J_1}_0}\otimes
L\right]^{J}}\right\rangle}^{\ast}, \label{w6}
\end{eqnarray}
\end{widetext}
where $A$ is the normalization factor and $J_{10}$ is a quantum
number similar to $J_0$.
Since this matrix element does not depend on $g_m$, $k_m$, $L_m$,
$Q_m$, $k'_m$, or $L'_m$, Eq. (\ref{w6}) can be further simplified
as
\begin{eqnarray}
|T|^2 &=&\sum_{J_0,J_{1_0}} \sum_{J'_0,{J'_1}_0}\mathcal{C}\left(g,k,L,J_0,{J_1}_0,J;k',Q,L',J'_0,{J'_1}_0,J'\right)\nonumber\\
&&\times\left\langle {\left[\left[{(c\bar c)_{g'}\otimes L}\right]_{k'} \otimes \left[Q\otimes L'\right]_{J'_0}\right]^{J}}{\parallel}H{\parallel}{\left[\left[{\bar{D}_g}^{(\ast)} {\Sigma_k}^{(\ast)}\right]_{J_0}\otimes L\right]^{J}} \right\rangle \nonumber\\
&&\times {\left\langle {\left[ \left[{(c\bar c)_{g'}\otimes L}\right]_{k'} \otimes
\left[Q\otimes
L'\right]_{{J'_1}_0}\right]^{J}}{\parallel}H{\parallel}{\left[\left[{\bar{D}_g}^{(\ast)}
{\Sigma_k}^{(\ast)}\right]_{{J_1}_0}\otimes
L\right]^{J}}\right\rangle}^{\ast}\nonumber\\\label{w7}
\end{eqnarray}
with
\begin{widetext}
\begin{eqnarray}
&&\mathcal{C}\left(g,k,L,J_0,J1_0,J;k',Q,L',J'_0,J1'_0\right)
\nonumber\\&&=A\sum_{g_m,k_m,{L_m}} \sum_{k'_m,Q_m,{L'_m}} \sum_{J_0,{J_1}_0} \sum_{J'_0,{J'_1}_0}{{\left\langle J'_0(k'_m+Q_m)|k'{k'_{m}},QQ_m\right\rangle}^{\ast}}   {\left\langle J{J_m}|J'_0(k'_m+Q_m)L'{L'_{m}}\right\rangle}^{\ast} \left\langle J_0(g_m+k_m)|g{g_m}k{k_{m}}\right\rangle  \nonumber\\
&&\quad\times  {\left\langle J(g_m+k_m+L_m)|J_0(g_m+k_m)L{L_{m}}\right\rangle}  {{\left\langle {J'_1}_0(k'_m+Q_m)|k'{k'_{m}},QQ_m\right\rangle} {\left\langle J{J_m}|{J'_1}_0(k'_m+Q_m)L'{L'_{m}}\right\rangle}}   \nonumber\\
&&\quad\times {{\left\langle
{{J_1}_0}(g_m+k_m)|g{g_m}k{k_{m}}\right\rangle}^{\ast} {\left\langle
J(g_m+k_m+L_m)|{{J_1}_0}(g_m+k_m)L{L_{m}}\right\rangle}^{\ast}}
{\delta}\left({g_m}+{k_m}+{L_m}-{k'_m}-{Q_m}-{L'_m}\right).\label{w8}
\end{eqnarray}
\end{widetext}
Through the derivation of Eqs. (\ref{w7}) and (\ref{w8}), we
prove that different $J_0$ ($J'_0$) contributions in the initial
(final) state for a fixed $J$ are equivalent when $J_0={J_1}_0$ and
$J'_0={J'_1}_0$. If $J_0\neq {J_1}_0$ or $J'_0\neq {J'_1}_0$, the
coefficient
$\mathcal{C}\left(g,k,L,J_0,J_{10},J;k',Q,L',J'_0,J'_{10}\right)$
vanishes, which shows that there is no mixing between different
$J_0$ and ${J_1}_0$ or $J'_0$ and ${J'_1}_0$ states. Now Eq. (\ref{w7}) becomes quite simple,
\begin{widetext}
\begin{eqnarray}
|T|^2 =\sum_{J_0} \sum_{J'_0}\frac{1}{(2 J_0+1)(2J'_0+1)}\left
|\left\langle {\left[{(c\bar c)}_{k'} \otimes \left[Q\otimes
L'\right]_{J'_0}\right]^{J}}{\parallel}H{\parallel}{\left[\left[{\bar{D}_g}^{(\ast)}
{\Sigma_k}^{(\ast)}\right]_{J_0}\otimes L\right]^{J}}
\right\rangle\right|^2 . \label{w9}
\end{eqnarray}
\end{widetext}

Combining Eqs. (\ref{ww1}) and (\ref{w2}) with Eq. (\ref{w9}) and
using the heavy quark symmetry, we can obtain useful information
about the $T$ matrix element for the hidden-charm decays.

\section{Numerical results}
\label{sec3}

With the preparation discussed in Sec. \ref{sec2}, we focus on the ratios of
the partial decay widths of the hidden-charm molecular pentaquarks
using the spin rearrangement scheme. The spin structure of the initial
state has been decomposed into heavy and light spins and labeled by
three quantum numbers, $h$, $R'$, and $T$, which are introduced in
Eq. (\ref{ww1}). For the final states, the total spin $R$ is fixed
as ${1}/{2}$ since we only consider the decay modes which are
involved with a nucleon and a charmonium.

Since the D-wave or higher partial waves are strongly suppressed by
phase space, we focus on the S-wave and P-wave decay of these
states. Throughout our calculation, we redefine the matrix element as
\begin{eqnarray}
H_{R,R'}=\left\langle
{[h,R,T]^{J}}{\parallel}H{\parallel}[h,R',T]]\right\rangle,
\end{eqnarray}
which corresponds on the matrix element to the right-hand side of
Eq. (\ref{w9}), where $|[h,R,T]\rangle$ and $|[h,R',T]\rangle$ are
the abbreviations of the initial and final states, respectively. In
Table \ref{tab1}, we collect the $H_{R,R'}$ related $T$ matrix
element, where the results depend on two unknown matrix elements, ${H_{\frac{1}{2},\frac{1}{2}}}$ and ${H_{\frac{1}{2},\frac{3}{2}}}$.
These matrix elements depend on the specific decay dynamics. Hence, they cannot be determined by the symmetry analysis alone.

\renewcommand{\arraystretch}{2}
\begin{table*}[htbp]
\begin{center}
{\scriptsize \caption{The relevant $H_{R,R'}$ matrix elements of the
hidden-charm molecular pentaquarks with $J^{P}={\frac{3}{2}}^{\pm}$
and ${\frac{5}{2}}^{\pm}$. Here, $\cdots$ indicates that the decay
channel is forbidden for the S-wave and P-wave decays.} \label{tab1}
   \begin{tabular}{c| c|ccccccccc} \toprule[1pt]
    $I(J^{P})$ & {Initial state} & \multicolumn{4}{c}{Final state} \\\midrule[1pt]
      & & ${J/{\psi}N}$ & $\chi_{c0}N$ & ${{\chi_{c1}}N}\,(J_0=\frac{1}{2})$  & ${{\chi_{c1}}N}\,(J_0=\frac{3}{2})$ \\
      \multirow{4}{*}{${\frac{1}{2}}({\frac{3}{2}}^{-})$} & ${\bar{D}}^{\ast}{{\Lambda_c}}$     &$H_{\frac{1}{2},\frac{1}{2}}$    & $\cdots$     &$\cdots$   &$\cdots$     \\

      &$\bar{D}{\Sigma^{\ast}}_c$   &   $-\frac{1}{\sqrt{3}}H_{\frac{1}{2},\frac{1}{2}}$    & $-\frac{1}{3\sqrt{2}}{H_{\frac{1}{2},\frac{1}{2}}}-\frac{\sqrt{5}}{6}{H_{\frac{1}{2},\frac{3}{2}}}$      &$\cdots$    &$\cdots$ \\

      &${\bar{D}}^{\ast}{{\Sigma_c}}$ &$\frac{1}{3}{H_{\frac{1}{2},\frac{1}{2}}}$    &$\frac{1}{3\sqrt{6}}{H_{\frac{1}{2},\frac{1}{2}}}-\frac{\sqrt{5}}{3\sqrt{3}}{H_{\frac{1}{2},\frac{3}{2}}}$    & $-\frac{1}{3\sqrt{6}}{H_{\frac{1}{2},\frac{1}{2}}}+\frac{5}{3\sqrt{6}}{H_{\frac{1}{2},\frac{3}{2}}}$   &  $-\frac{\sqrt{5}}{6\sqrt{3}}{H_{\frac{1}{2},\frac{1}{2}}}+\frac{\sqrt 2}{3\sqrt{3}}{H_{\frac{1}{2},\frac{3}{2}}}$   \\

      &${\bar{D}}^{\ast}{{\Sigma_c}^{\ast}}$ & $\frac{\sqrt 5}{3}{H_{\frac{1}{2},\frac{1}{2}}}$    &$\frac{\sqrt 5}{3\sqrt 6}{H_{\frac{1}{2},\frac{1}{2}}}-\frac{1}{6\sqrt 3}{H_{\frac{1}{2},\frac{3}{2}}}$
      & $-\frac{\sqrt 5}{3\sqrt 6}{H_{\frac{1}{2},\frac{1}{2}}}+\frac{\sqrt 5}{6\sqrt 6}{H_{\frac{1}{2},\frac{3}{2}}}$
     & $-\frac{5}{6\sqrt 3}{H_{\frac{1}{2},\frac{1}{2}}}+\frac{1}{6}\sqrt{\frac{2}{15}}{H_{\frac{1}{2},\frac{3}{2}}}$      \\
      \midrule[1pt]

      & & ${{\chi_{c2}}N}\,(J_0=\frac{1}{2})$
      & ${{\chi_{c2}}N}\,({J_0}=\frac{3}{2})$ & ${h_{c}(1p)N}\,({J_0}=\frac{1}{2})$ & ${h_{c}(1p)N}\,({J_0}=\frac{3}{2})$  \\
      \multirow{4}{*}{${\frac{1}{2}}({\frac{3}{2}}^{-})$} & ${\bar{D}}^{\ast}{{\Lambda_c}}$   &$\cdots$    & $\cdots$   &$\cdots$    & $\cdots$     \\

      &$\bar{D}{\Sigma^{\ast}}_c$  &$\cdots$    & $\cdots$  &  $\cdots$    & $\cdots$      \\

      &${\bar{D}}^{\ast}{{\Sigma_c}}$  &$\cdots$    & $\cdots$    &$\cdots$     &$\cdots$\\

      &${\bar{D}}^{\ast}{{\Sigma_c}^{\ast}}$ & $\frac{5}{3\sqrt 6}{H_{\frac{1}{2},\frac{1}{2}}}+\frac{1}{6\sqrt 6}{H_{\frac{1}{2},\frac{3}{2}}}$   & $\frac{5}{6\sqrt 3}{H_{\frac{1}{2},\frac{1}{2}}}+\frac{1}{3}{\sqrt{\frac{2}{15}}}{H_{\frac{1}{2},\frac{3}{2}}}$ & $\frac{\sqrt 5}{2\sqrt 3}{H_{\frac{1}{2},\frac{3}{2}}}$   & $\frac{\sqrt 5}{2\sqrt 3}{H_{\frac{1}{2},\frac{3}{2}}}$   \\
 \midrule[1pt]

      & & ${{\eta}_{c}N}$& ${J/{\psi}N}\,({J_0}=\frac{1}{2})$ & ${J/{\psi}N}\,({J_0}=\frac{3}{2})$ & ${{\chi_{c1}}N}$    \\
      \multirow{9}{*}{${\frac{1}{2}}({\frac{3}{2}}^+)$} & ${\bar{D}}^{\ast}{{\Lambda_c}}\,({J_0}=\frac{1}{2})$     &$\frac{\sqrt3}{2}{H_{\frac{1}{2},\frac{1}{2}}}$    & $\frac{1}{3}{H_{\frac{1}{2},\frac{1}{2}}}$     &$-\frac{\sqrt5}{6}{H_{\frac{1}{2},\frac{1}{2}}}$     &$\cdots$  \\

& ${\bar{D}}^{\ast}{{\Lambda_c}}\,({J_0}=\frac{3}{2})$     &0    & $\frac{\sqrt 5}{3}{H_{\frac{1}{2},\frac{1}{2}}}$     &$\frac{2}{3}{H_{\frac{1}{2},\frac{1}{2}}}$     &$\cdots$ \\
      &$\bar{D}{\Sigma^{\ast}}_c\,(J_0=\frac{3}{2})$   &   $\frac{1}{2}{H_{\frac{1}{2},\frac{3}{2}}}$    & $-\frac{1}{3}\sqrt{\frac{5}{3}}{H_{\frac{1}{2},\frac{1}{2}}}-\frac{1}{3}\sqrt{\frac{5}{6}}{H_{\frac{1}{2},\frac{3}{2}}}$    & $-\frac{2}{3}\sqrt{\frac{1}{3}}{H_{\frac{1}{2},\frac{1}{2}}}+\frac{11}{6}\sqrt{\frac{1}{15}}{H_{\frac{1}{2},\frac{3}{2}}}$   & $\cdots$     \\

      &${\bar{D}}^{\ast}{{\Sigma_c}}\,({J_0}=\frac{1}{2})$ &$-\frac{1}{2\sqrt{3}}{H_{\frac{1}{2},\frac{1}{2}}}$    &$-\frac{5}{9}{H_{\frac{1}{2},\frac{1}{2}}}+\frac{1}{9\sqrt{2}}{H_{\frac{1}{2},\frac{3}{2}}}$    &$\frac{5\sqrt 5}{18}{H_{\frac{1}{2},\frac{1}{2}}}-\frac{2}{9}{H_{\frac{1}{2},\frac{3}{2}}}$  &$\frac{35}{18\sqrt{{6}}}{H_{\frac{1}{2},\frac{1}{2}}}+(-\frac{1}{18\sqrt{3}}-\frac{2\sqrt{5}}{9\sqrt{6}}){H_{\frac{1}{2},\frac{3}{2}}}$   \\

&${\bar{D}}^{\ast}{{\Sigma_c}}\,({J_0}=\frac{3}{2})$ &$-\frac{1}{\sqrt{3}}{H_{\frac{1}{2},\frac{3}{2}}}$    &$\frac{\sqrt 5}{9}{H_{\frac{1}{2},\frac{1}{2}}}-\frac{\sqrt 10}{9}{H_{\frac{1}{2},\frac{3}{2}}}$    &$\frac{2}{9}{H_{\frac{1}{2},\frac{1}{2}}}+\frac{11}{9 \sqrt5}{H_{\frac{1}{2},\frac{3}{2}}}$  &$\frac{\sqrt5}{9\sqrt{{6}}}{H_{\frac{1}{2},\frac{1}{2}}}+(\frac{\sqrt10}{9\sqrt{6}}+\frac{11}{9\sqrt{6}}){H_{\frac{1}{2},\frac{3}{2}}}$    \\

      &${\bar{D}}^{\ast}{{\Sigma_c}^{\ast}}\,({J_0}=\frac{1}{2})$ & $\sqrt{\frac{2}{3}}{H_{\frac{1}{2},\frac{1}{2}}}$
&$-\frac{2 \sqrt2}{9}{H_{\frac{1}{2},\frac{1}{2}}}-\frac{1}{18}{H_{\frac{1}{2},\frac{3}{2}}}$  &$\frac{\sqrt 10}{9}{H_{\frac{1}{2},\frac{1}{2}}}+\frac{\sqrt 2}{9}{H_{\frac{1}{2},\frac{3}{2}}}$      &$\frac{7}{9\sqrt{{3}}}{H_{\frac{1}{2},\frac{1}{2}}}+(\frac{1}{18\sqrt{6}}+\frac{\sqrt 5}{9\sqrt{3}}){H_{\frac{1}{2},\frac{3}{2}}}$     \\

 &${\bar{D}}^{\ast}{{\Sigma_c}^{\ast}}\,({J_0}=\frac{3}{2})$ & $\frac{1}{2}\sqrt{\frac{5}{3}}{H_{\frac{1}{2},\frac{3}{2}}}$
&$\frac{5}{9}{H_{\frac{1}{2},\frac{1}{2}}}-\frac{1}{9\sqrt 2}{H_{\frac{1}{2},\frac{3}{2}}}$  &$\frac{2\sqrt 5}{9}{H_{\frac{1}{2},\frac{1}{2}}}+\frac{11}{90}{H_{\frac{1}{2},\frac{3}{2}}}$      &$\frac{5}{9\sqrt{{6}}}{H_{\frac{1}{2},\frac{1}{2}}}+(\frac{1}{18\sqrt{3}}+\frac{11\sqrt 5}{90\sqrt{6}}){H_{\frac{1}{2},\frac{3}{2}}}$     \\

&${\bar{D}}^{\ast}{{\Sigma_c}^{\ast}}\,({J_0}=\frac{5}{2})$ & $0$
&$\frac{\sqrt 3}{2}{H_{\frac{1}{2},\frac{3}{2}}}$  &$\frac{\sqrt 6}{5}{H_{\frac{1}{2},\frac{3}{2}}}$      &$(-\frac{1}{2\sqrt{{2}}}+\frac{1}{\sqrt 5}){H_{\frac{1}{2},\frac{1}{3}}}$       \\

      \midrule[1pt]

      &  & ${{\chi_{c2}}N}$  & ${h_{c}(1p)N}$  \\
      \multirow{3}{*}{${\frac{1}{2}}^+({\frac{3}{2}}^{+})$}  &${\bar{D}}^{\ast}{{\Sigma_c}^{\ast}}\,({J_0}=\frac{1}{2})$   &$-\frac{\sqrt 5}{9\sqrt{{3}}}{H_{\frac{1}{2},\frac{1}{2}}}+(-\frac{\sqrt 5}{18\sqrt{6}}+\frac{1}{9\sqrt{3}}){H_{\frac{1}{2},\frac{3}{2}}}$  &$\sqrt{\frac{3}{2}} {H_{\frac{1}{2},\frac{1}{2}}}$   \\

 &${\bar{D}}^{\ast}{{\Sigma_c}^{\ast}}\,({J_0}=\frac{3}{2})$      &$\frac{7\sqrt 5}{9\sqrt{{6}}}{H_{\frac{1}{2},\frac{1}{2}}}+(\frac{11}{90\sqrt{6}}-\frac{\sqrt 5}{18\sqrt{3}}){H_{\frac{1}{2},\frac{3}{2}}}$  &$\frac{\sqrt{5}}{2\sqrt{3}}H_{\frac{1}{2},\frac{3}{2}}$ \\

&${\bar{D}}^{\ast}{{\Sigma_c}^{\ast}}\,({J_0}=\frac{5}{2})$    &$(\frac{\sqrt 5}{2\sqrt{{2}}}+\frac{1}{5}){H_{\frac{1}{2},\frac{1}{3}}}$  &$0$ \\

           \midrule[1pt]

      \multirow{1}{*}
      & & ${J/{\psi}N}$ & $\chi_{c0}N$ & ${{\chi_{c1}}N}(J_0=\frac{3}{2})$ & ${{\chi_{c2}}N}\,(J_0=\frac{1}{2})$ \\
      {${\frac{1}{2}}({\frac{5}{2}}^{-})$} &${\bar{D}}^{\ast}{{\Sigma_c}^{\ast}}$  &$\cdots$    &$\cdots$       &$-\sqrt{\frac{3}{10}}{H_{\frac{1}{2},\frac{3}{2}}}$   & ${H_{\frac{1}{2},\frac{3}{2}}}$ \\

\midrule[1pt]

      \multirow{1}{*}
      & &  ${{\chi_{c2}}N}\,(J_0=\frac{3}{2})$ & ${h_{c}(1p)N}(J_0=\frac{3}{2})$  & ${{\eta}_{c}N}$\\
   {${\frac{1}{2}}({\frac{5}{2}}^{-})$} &${\bar{D}}^{\ast}{{\Sigma_c}^{\ast}}$ & $\sqrt{\frac{7}{10}}{H_{\frac{1}{2},\frac{3}{2}}}$  &0 &$\cdots$ \\

\midrule[1pt]

      & & $J/{\psi}N$ & $\chi_{c2}N$ & \\
      \multirow{6}{*}{${\frac{1}{2}}({\frac{5}{2}}^{+})$}   &${\bar{D}^{\ast}{\Lambda}_c}\,(J_0=\frac{3}{2})$   &   ${H_{\frac{1}{2},\frac{3}{2}}}$       & $\cdots$   \\
      &${\bar{D}{\Sigma^{\ast}}_c}\,(J_0=\frac{3}{2})$   &   $-\frac{1}{\sqrt 3}{H_{\frac{1}{2},\frac{1}{2}}}-\sqrt{\frac{1}{15}}{H_{\frac{1}{2},\frac{3}{2}}}$       & $\cdots$   \\

      &${{\bar{D}}^{\ast}{{\Sigma_c}}}\,(J_0=\frac{3}{2})$ & $\frac{1}{3}{H_{\frac{1}{2},\frac{1}{2}}}-{\frac{2}{3\sqrt{5}}}{H_{\frac{1}{2},\frac{3}{2}}}$      &$\cdots$   \\

      &${\bar{D}}^{\ast}{{\Sigma_c}^{\ast}}\,(J_0=\frac{3}{2})$ & $\frac{\sqrt5}{3}{H_{\frac{1}{2},\frac{1}{2}}}-{\frac{1}{15}}{H_{\frac{1}{2},\frac{3}{2}}}$    &$\frac{1}{10}\sqrt{\frac{7}{3}}{H_{\frac{1}{2},\frac{3}{2}}}$      \\

  &${\bar{D}}^{\ast}{{\Sigma_c}^{\ast}}\,(J_0=\frac{5}{2})$ & $\frac{\sqrt21}{5}{H_{\frac{1}{2},\frac{3}{2}}}$    &$\frac{2}{5}{H_{\frac{1}{2},\frac{3}{2}}}$      \\

      \bottomrule[1pt]

      \end{tabular}}
\end{center}
\end{table*}



In Table \ref{tab1}, we first illustrate the decay patterns of the
$J^P={{3}/{2}}^{-}$ hidden-charm molecular pentaquarks composed of
$(\bar{D}\Sigma_c^*)$, $(\bar{D}^*\Sigma_c)$, and
$(\bar{D}^*\Sigma_c^*)$, respectively. In the heavy quark symmetry
limit, the $D$ and $D^*$ mesons have the same spatial wave function.
Similarly, $\Sigma_c$ and $\Sigma_c^*$ have the same spatial wave
function. For the $J/\psi N$ decay mode of the above three types of
molecular states, we notice that the decay process depends on one
common matrix element $H_{{\frac{1}{2}},\frac{1}{2}}$. Therefore, we
get the ratio
\begin{eqnarray}
\Gamma\left[(\bar D{\Sigma^{\ast}_c})\right]:\Gamma\left[({\bar
D}^{\ast}{\Sigma}_c)\right]:\Gamma\left[({\bar
D}^{\ast}{\Sigma^{\ast}_c})\right]=2.7:1.0:5.4,
\end{eqnarray}
where we have included the contribution of the corresponding phase
spaces. Compared with the $J/\psi N$ decay mode, the other P-wave
decay modes, like $\chi_{c0}N$, $\chi_{c1}N\,(J_0=\frac{1}{2})$, and
$\chi_{c1}N\,(J_0=\frac{3}{2})$, depend on two unknown matrix
elements. Moreover, they are strongly suppressed by the limited
phase space by the factor $|\mathbf {p}_N|^3$, where $\mathbf{p}_N$
is the decay momentum of the nucleon.

For the hidden-charm molecular pentaquarks with $J^P={{3}/{2}}^{+}$,
their decay pattern is quite different than that of the
$J^P={{3}/{2}}^{-}$ states. In Table \ref{tab1}, we use $0$ to
denote the decay channels which are suppressed due to the heavy
quark symmetry. This feature reflects the inner structure of the
pentaquark state with $J^P={{3}/{2}}^{+}$. For these hidden-charm
molecular pentaquarks with the same $I(J^P)$ quantum number and the
same configuration, the corresponding matrix elements are still
different if the corresponding $J_0$ quantum number is different, as
shown in Table \ref{tab1}. For the $\eta_c N$ decay mode, the ratio
of the partial decay widths of the
$({\bar{D}}^{\ast}{{\Sigma_c}})\,({J_0}=\frac{1}{2})$ and
$({\bar{D}}^{\ast}{\Sigma_c^{\ast}})\,({J_0}=\frac{1}{2})$ molecular
states is
\begin{eqnarray}
\Gamma\left[({\bar{D}}^{\ast}{{\Sigma_c}})\,({J_0}=\frac{1}{2})\right]:\Gamma\left[({\bar{D}}^{\ast}{\Sigma_c^{\ast}})\,({J_0}=\frac{1}{2})\right]=1:9.7,
\end{eqnarray}
and the ratio of the partial decay widths of the
$({\bar{D}}{\Sigma_c^{\ast}})\,({J_0}=\frac{3}{2})$,
$({\bar{D}}^{\ast}{{\Sigma_c}})\,({J_0}=\frac{3}{2})$, and
$({\bar{D}}^{\ast}{\Sigma_c^{\ast}})\,({J_0}=\frac{3}{2})$ states is
{\small
\begin{eqnarray}
&&\Gamma\left[({\bar{D}}{\Sigma_c^{\ast}})\,({J_0}=\frac{3}{2})\right]:\Gamma\left[({\bar{D}}^{\ast}{{\Sigma_c}})\,({J_0}=\frac{3}{2})\right]:\Gamma\left[({\bar{D}}^{\ast}{\Sigma_c^{\ast}})\,({J_0}=\frac{3}{2})\right]\nonumber\\&&=1.0:1.7:2.6,
\end{eqnarray}}

For the S-wave ${\bar{D}}^{\ast}{\Sigma_c^{\ast}}$ states with
$J^P={{5}/{2}}^-$, its $h_{c}(1P)N$ decay channel is strongly
suppressed in the heavy quark symmetry. The ratio of the partial
decay widths of its $\chi_{c1}(1P)N$,
$\chi_{c2}(1P)N\,(J_0=\frac{1}{2})$, and
$\chi_{c2}(1P)N\,(J_0=\frac{3}{2})$ modes reads
{\small
\begin{eqnarray}
&&\Gamma\left[{{\chi_{c1}}(1P)N}\right]:\Gamma\left[{{\chi_{c2}}(1P)N}\,(J_0=\frac{1}{2})\right]:\Gamma\left[{{\chi_{c2}}(1P)N}\,(J_0=\frac{3}{2})\right]\nonumber\\&&=1.5:1.4:1.0,
\end{eqnarray}}
where we have also considered the phase space correction.
Unfortunately, for these states with $J^P=\frac{5}{2}^+$, there does
not exist any simple relation.

\section{Summary}

\label{sec4}

Inspired by the recent experimental observation of the two $P_c$
states \cite{Aaij:2015tga}, we have studied the decay behaviors of
the hidden-charm pentaquarks with the $\bar{D}^{(*)}\Sigma_c^{(*)}$
configuration and spin-parity $J^P=3/2^\pm, 5/2^\pm$ through the
spin rearrangement scheme. Within this framework, we have obtained
several ratios of the partial decay widths of some decay channels of
the hidden-charm pentaquarks, which will be useful in the further
experimental investigation of their inner structures.

We also notice that most of the partial decay widths of the
hidden-charm pentaquarks depend on the unknown matrix element
$H_{R,R'}$, which is governed by strong decay dynamics. Such matrix
elements can either be calculated with a phenomenological model or
extracted through the experimental measurement of several partial
decay widths. Once they are known, all of the other hidden-charm decay
widths can be predicted.

The two observed $P_c$ states open a Pandora's box of exotic state
studies. In the coming years, more and more novel phenomena are
expected with experimental progress, especially from LHCb and the
forthcoming BelleII. Clearly, more theoretical explorations of the
hidden-charm pentaquark states are desirable.

\subsection*{Acknowledgments}
This project is supported by the National Natural Science Foundation
of China under Grants No. 11205011, No. 11475015, No. 11375024, No.
11222547, No. 11175073, No. 11035006, and No. 11575008. \vfil

\end{document}